\documentstyle[12pt]{article}

\def\ba{\begin{array}}
\def\ea{\end{array}}
\def\be{\begin{equation}}
\def\ee{
\end{equation}}
\def\bea{\begin{equation}\begin{array}{l}}
\def\eea{\end{array}\end{equation}}
\def\f{\frac}
\def\om{\omega}
\def\Om{\Omega}
\def\we{\wedge}
\def\d{\delta}

\def\ep{\epsilon}

\def\c{\cite}

\def\n{\nabla}
\def\a{\alpha}
\def\b{\beta}
\def\pa{\partial}
\begin{document}\begin{flushright}
AS-ITP-2001-009\\
\vspace{1ex}
 April 16, 2001
\end{flushright}
\vskip 2mm
\vskip 10mm
\begin{center}
{\large \bf
SYMPLECTIC, MULTISYMPLECTIC STRUCTURES \\
AND\\
\vskip 1mm

 EULER-LAGRANGE COHOMOLOGY}

\vspace{10mm}

H.Y. Guo  
, \quad Y.Q. Li, 
 \quad K. Wu{\footnote{Emails: hyguo@itp.ac.cn; qylee@itp.ac.cn; wuke@itp.ac.cn}}\\
Institute of Theoretical    Physics,    Academia   Sinica\\
P.O. Box 2735, Beijing 100080, China.

\vskip 2mm
and\\
\vskip 2mm
S.K. Wang{\footnote{Email: xyswsk@pku.edu.cn}}\\
Institute of Applied mathematics\\
Academy of Mathematics and Systems Sciences, Academia Sinica\\
P.O. Pox 2734, Beijing 100080, China. 
\end{center}

\vskip 10mm\vskip 15mm

\noindent \centerline {\sc Abstract} \vskip 2mm
We study the Euler-Lagrange cohomology  and explore the symplectic
or multisymplectic geometry and their preserving properties in
classical mechanism and classical field theory in
Lagrangian and Hamiltonian formalism in each case respectively. By
virtue of the Euler-Lagrange cohomology that is nontrivial in the
configuration space, we show that the symplectic or
multisymplectic geometry and related preserving property can be
established not only in the solution space but also in the
function space if and only if the relevant closed Euler-Lagrange
cohomological condition is satisfied in each case. We also apply
the cohomological approach directly to Hamiltonian-like ODEs and
Hamiltonian-like PDEs no matter whether there exist known
Lagrangian and/or Hamiltonian associated with them.
\vskip 5mm

\newpage
{\parindent=0mm {\sc Contents} \it \vskip 3mm

 1. Introduction

2. Euler-Lagrange Cohomology and Symplectic Structure Preserving
in Classical Mechanism

2.1. Euler-Lagrange cohomology and symplectic structure preserving
in Lagrangian mechanism

2.2  Euler-Lagrange cohomology and symplectic structure preserving
in Hamiltonian formalism

3. Euler-Lagrange Cohomology and Multisymplectic Structure
Preserving in Classical Field Theory

3.1 Euler-Lagrange cohomology and multisymplectic structure
preserving for classical field theory in Lagrangian formalism

3.2 Euler-Lagrange cohomology and multisymplectic structure
preserving for classical field theory in Hamiltonian formalism

4 Cohomological Approach to  Hamiltonian-like ODEs, PDEs and Their
Symplectic and Multisymplectic Properties

4.1 Symplectic cohomological approach to  Hamiltonian-like ODEs
and their symplectic structure preserving law

4.2 Symplectic cohomological approach to  Hamiltonian-like PDEs
and their multisymplectic structure preserving law

5. Remarks

Acknowledgements

References }
\section{Introduction}

$\quad $ It is well known that the symplectic and multisymplectic
structures play  important roles in the classical Hamiltonian
mechanism \c{ALD78} \c{AM78} as well as in the classical
Lagrangian and/or Hamiltonian field theories \c{BSF88} \c{TB97}
\c{GIM97} \c{MS98} \c{MPS98}  respectively. For a quite long
period,
 in the usual
standard approach, the symplectic or multisymplectic structure and their preserving
 properties have always been established in the solution space of the equation(s) of motion in
each case. Very recently, however, it has been found that the
symplectic structure and its preserving law in classical mechanics
or the multisymplectic structure and its preserving  law in
classical scalar field theory hold not only in the solution space
but also in the function space characterized by what has been
named the Euler-Lagrange  cohomology in each case \c{GLW01}
\c{hyg1}. The Euler-Lagrange cohomology, in fact,  not only
provides some  powerful concepts and method to study the
symplectic or multisymplectic structures and their preserving
properties in the classical mechanism and classical field theory
but also opens up an important  new research subject for further
investigation. In fact,  this Euler-Lagrange cohomology approach
have already been generalized to the difference discrete cases and
applied to the symplectic, multisymplectic algorithms \c{GLW01}
\c{hyg1} \c {GLW02} as well as the simple finite element method to
find some symplectic or multisymplectic structure preserving
properties in certain schemes derived from the finite element
method \c{GJLW03} \c{kw1}.


In the approach in \c{GLW01} \c{hyg1}, the variation of the action
functional had been dealt with in the
traditional way in physics. 
However, the variation of the action functional may be treated as
the derivative
with respect to the free parameters as long as  the free parameter
space associated with the variation of the variables is introduced
in the configuration space.  As a matter of fact, it is quite
common in some above-mentioned standard approaches to  deal with
the variation of given functional as the derivative with respect
to one free parameter associated with the variation of the
variables in the configuration space. 
 Instead of introducing one free parameter in some standard
approaches, we introduce a family of free parameters to  describe
the variation of the variables in all directions of the
configuration space.

We  also employ the conception of exterior differential
calculus in the general function space ( see, for example, \c
{AC94} ) as what have been implicated in \c{GLW01} \c{hyg1}. On the
other hand, however, in order to more simply comprehend the
exterior differential calculus in the general function space,  we
present in a simpler and more manifest way how to manipulate it by
introducing an associated free parameter space with the functions
at each point of base manifold and regarding the exterior
differential calculus of the functions at this point 
as the differential calculus with respect to the free
parameter space.

Therefore, the above two aspects, namely, the
variation of the action functional etc and the exterior
differentiation of the Lagrangian, Hamiltonian etc may be treated
in a same framework as the first derivative and the second (exterior)
derivative with respect to the free parameters as long as  the
parameter space associated with the variables at each point of the
base manifold is introduced in the configuration space.

In this paper, we establish further in  more rigorous and  more evident way the
geometric foundation for some relevant issues in \c{GLW01}\c{hyg1}, such
as the Euler-Lagrange cohomology,
 its relation with the symplectic structure and its preserving property as well as
 its relation with the
multisymplectic structure and its preserving property
 in classical mechanics and classical  field theory for a set of generic fields in the
 Lagrangian
formalism respectively. In addition, we establish in the
Hamiltonian formalism the Euler-Lagrange cohomology, symplectic or
multisymplectic structure and relevant conservation law for the
classical mechanism and classical field theory for a set of
generic fields. Furthermore, we also apply the Euler-Lagrange-like
cohomological approach directly to certain types of ODEs and PDEs
that may be called the Hamiltonian-like ODEs and the
Hamiltonian-like PDEs no matter whether there exist the associated
Lagrangian and/or Hamiltonian at all. We find that the
cohomological scenario is available to so-called the Hamiltonian
PDEs introduced in \c{TB97}. In fact, it is  a type of
Hamiltonian-like PDEs in the terminology of this paper.

The plan of this paper is as follows. We first  re-derive some
well-known content on symplectic structure and its preserving
property for classical mechanism in Lagrangian formalism and
Hamiltonian formalism respectively in section 2. The important
issues of this section are about the free parameter differential
calculus approach to
 the variation of the functional  and to the exterior
differential calculus of the functional in the generic function
space, as well as the introduction of what has been called the
Euler-Lagrange cohomology and relevant concepts and content in
\c{GLW01} \c{hyg1} such as the Euler-Lagrange 1-forms, the
coboundary Euler-Lagrange 1-forms, the Euler-Lagrange conditions
and their role-played in order to explain those symplectic
geometry and relevant preserving property for the classical
mechanism in both the Lagrangian and Hamiltonian formalism.  In
the subsection 2.1. we first deal with the Lagrangian mechanism.
Then  in the subsection 2.2, we transfer to the Hamiltonian
formalism. We show that the relevant Euler-Lagrange cohomology in
each case is nontrivial and it is directly linked with the
symplectic structure preserving law in classical mechanism.  As a
matter of fact, similar to the existence of symplectic structure,
the symplectic structure preserving law in classical mechanism
holds not only in the solution space of the Euler-Lagrange
equation in Lagrangian formalism or the canonical equations in
Hamiltonian formalism but also in the function space associated
with the Euler-Lagrange cohomology in general. In section 3, we
study  the multisymplectic structure and its preserving property
in classical field theory for a set of generic fields in the
Minkowskian spacetime in both Lagrangian In the subsection 3.1, we
first deal with the variation of the action functional as the
differentiation with respect to multi-parameter in the
configuration space. We also employ multi-parameter differential
calculus approach to the exterior derivative in the generic
function space to introduce the Euler-Lagrange cohomology in the
Lagrangian formalism. We show that the nontriviality of the
cohomology is directly linked with the nontriviality of the
canonical 1-forms. We also prove that the necessary and sufficient
condition for the multisymplectic structure and its preserving
property, i.e. the conservation law, is the relevant
Euler-Lagrange form being closed. Then in the subsection 3.2, we
transfer to the Hamiltonian formalism by introducing the canonical
conjugate field variables and making Legendre transformation. We
also show that the nontriviality of the Euler-Lagrange cohomology
and its relation with the multisymplectic structure and its
conservation law in the Hamiltonian formalism. In the section 4,
we apply the cohomological scenario directly to what we call
certain types of Hamiltonian-like ODEs and of Hamiltonian-like
PDEs. We  show in the subsection 4.1 that the Euler-Lagrange-like
cohomological scenario can be applied to certain types of
Hamiltonian-like ODEs no matter whether there exist known
associated Lagrangian and/or Hamiltonian. In the subsection 4.2,
we emphasize that the Euler-Lagrange-like cohomological scenario
can also be directly applied to certain types of Hamiltonian-like
PDEs  no matter whether there exist known associated Lagrangian
and/or Hamiltonian for the systems. We show that the so-called
Hamiltonian PDEs proposed by Bridges in \c{TB97} are just the ones
in the type of Hamiltonian-like PDEs.
 For all these cases, the symplectic,  multisymplectic
structure and relevant conservation law  not only exist in the
solution space of the relevant equation(s) but also hold in the
function space associated with the relevant  cohomological
condition. Finally, we  end with some remarks in section 5.

\section{ Euler-Lagrange
Cohomology and Symplectic Structure Preserving
in Classical Mechanism}

$\quad $ In this section, we  recall some well-known content on
symplectic  structure and its preserving property in the
Lagrangian and Hamiltonian formalism for the classical mechanism. 
The important point is to introduce the Euler-Lagrange cohomology
associated with the Lagrangian functions and the Euler-Lagrange
equation in
Lagrangian formalism and the one in the Hamiltonian formalism. We 
 explain their
important roles in the symplectic
structure and its preserving property in these formalisms.

\subsection{ Euler-Lagrange
cohomology and symplectic structure preserving  in Lagrangian mechanism}

 $\quad $
 We begin with the Lagrangian mechanism.
 Let time $t\in R^1$ be the base manifold, $M$ the $n$-dimensional configuration space
 on $t$ with coordinates $q^i(t)$, $(i=1, \cdots, n)$,
$TM$ the tangent bundle of $M$ with coordinates $(q^i, \dot q^j)$,
where $\dot q^j$ is the derivative of $q^j$ with respect to $t$,
  $F(TM)$ the function space on $TM$.

The Lagrangian of the systems is denoted by $L(q^i, {\dot q^j})$.
The action functional along a curve $q(t)$ in $M$ can be
constructed by integrating of $L$ along the tangent of the curve
\be\label{S1} S(q(t)):=\int_a^b L(q^i(t), {\dot q^i(t)})dt. \ee

Let us consider the case that both $q^i$ and ${\dot q^j}$ variate
by an infinitesimal increments such that the curve becomes a
congruence of curves.
Then at each moment of $t$  in the congruence we can introduce the
infinitesimal variations of $q^i$ and ${\dot q^j}$ as follows
\be
q_{\ep} ^i(t)=q^i(t)+{\ep^k} \d q_k^i(t), \quad {\dot
q_{\ep}^j}=\dot q^j+{\ep^k} {\d {\dot q_k^j}},
\ee where ${\ep^k}$
are $n$ free parameters  along the $k$-th direction in the configuration
space, $\d q_{\ep^ k} ^i(t)$ and $\d {\dot q_{\ep^ k} ^j(t)}$
infinitesimal increments
 of $q_{\ep} ^i(t)$  and ${\dot q_{\ep} ^j}(t)$ at the moment $t$ along  the $k$-th direction in the
congruence of curves:
\be
 \d q_{\ep^k}^i(t):= \f d {d\ep^k} \mid_{\ep^k=0} q_{\ep}^i(t)=\d q_k^i(t), \qquad
\d {\dot q_{\ep^ k}^j}(t):=\f d {d\ep^k} \mid_{\ep^k=0} {\dot
q_{\ep}^j}(t)={\d {\dot q_k^j}}(t). \ee While the exterior
derivatives of $q_{\ep} ^i(t)$  and ${\dot q_{\ep} ^j}(t)$ at the
moment $t$ in the congruence of curves are given by: \be d q_{\ep
}^i:=\f {\pa q_{\ep}^i} {\pa {\ep^l} }  d{\ep^l} =d{\ep}^k \d
q_k^i,\qquad d {\dot q_{\ep }^j}:=\f {\pa \dot q_{\ep}^i} {\pa
{\ep^l} }  d{\ep^l} =d{\ep}^k \d \dot q_k^i. \ee

In the congruence of curves, the Lagrangian also becomes a family
of Lagrangian and the same for the action functional: \be S(q(t))
\rightarrow S_{\ep} ((q_{\ep}(t)))=\int_a^b L_{\ep}(q_{\ep}^i(t),
{\dot q_{\ep}^i(t)})dt, \ee where the upper-index $k$ of $\ep^k$
is omitted.

Hamilton's principle,  i.e. the least variational principle, seeks
the curve along which the action $S$ is stationary under
variations of $q^i(t)$ with fixed endpoints. This can be
manipulated by taking differentiation with respect to $\ep^k$  and
setting ${\ep^k}=0 $ afterwards: \be \d S(q(t)) :=\f d {d\ep^k}
\mid_{\ep^k=0} S_{\ep} (q_{\ep}(t)) =0 \ee for all $\d q_{\ep^k
}^i(t)=\d {q_k^i(t)} $  with $\d q_k^i(a) = \d q_k^i(b)=0$.

It is simple to get the differentiation  of the action with respect to ${\ep^k}$
\be
\label{dS1}
d S_{\ep} (q_{\ep}(t))=\f {\pa} {\pa {\ep^k}}
\int_a^b (L(q_{\ep}^i(t), {\dot q_{\ep}^i(t)})dt) d{\ep^k}
=\int_a^bdq_{\ep}^i \{\frac {\pa L_{\ep} }{\pa {q_{\ep} ^i}}-\frac
{d} {dt} \frac {\pa L_{\ep} } {\pa {\dot q_{\ep} ^i}}\} dt+\f {\pa
L_{\ep} } {\pa  \dot q_{\ep} ^j} dq_{\ep}^j \mid_a^b. \ee
Therefore, the variation of the action is given by \be \d S_{\ep}
(t)=\int_a^b \d q_{{\ep}^ k}^i \{\frac {\pa L}{\pa {q^i}}-\frac
{d} {dt} \frac {\pa L} {\pa {\dot q^i}}\} dt+\f {\pa L} {\pa  \dot
q^j} \d q_{{\ep}^ k}^j\mid_a^b. \ee The last term in the above
equation vanishes due to $\d q_k^i(a) = \d q_k^i(b)=0$, so that
the stationary requirement for $S$ yields the Euler-Lagrange
equations \be\label{ee1} \frac {\pa L}{\pa {q^i}}-\frac {d} {dt}
\frac {\pa L} {\pa {\dot q^i}}=0. \ee

It is obvious but important to see that in the above manipulation
the variation of the action functional, $\d S_{\ep} $, is very
closely linked with the differentiation of $S_{\ep} $ with respect
to the free parameters $\ep^k$ in the congruence, i.e.  $dS_{\ep}
$. Notice that the integrant in ({\ref{dS1}}) for $d S_{\ep} $ and
the boundary term are 1-forms with respect to $d\ep^k$.
 In addition, the
differentiation with respect to $\ep^k$ may be viewed as an
exterior derivative so that an exterior derivative operator $d$
that is nilpotent with respect to $\ep^k$ in the congruence may be
introduced.

The differentiation in the function space calculated for the
action is in fact completely relied on the derivative of  the
family of Lagrangian functions \be \label{lep1}
L_{\ep}:=L(q_{\ep}^i(t), {\dot q_{\ep}^i(t)}), \ee with respect to
$\ep^k$. Therefore, it follows that \be dL_{\ep}\mid_{\ep^k=0}
=\{\frac {\pa L}{\pa {q^i}}-\frac {d} {dt} \frac {\pa L} {\pa
{\dot q^i}}\} dq_{\ep}^i +\frac {d} {dt} \{\frac {\pa L} {\pa
{\dot q^i} } dq_{\ep}^i\}. \ee

Let us from now on denote
\be
dq^i:=dq_{\ep}^i,
\ee
and define the Euler-Lagrange  1-form and the canonical 1-form
 $\theta$ on $T^*M$,
\be
\label{E1}
E(q^i, \dot q^i):=\{\frac {\pa L}{\pa {q^i}}-\frac {d} {dt} \frac {\pa L} {\pa {\dot q^i}}\}
dq^i,
\ee
\be
\label {t1}
\theta = \frac {\pa L} {\pa {\dot q^i} } dq^i,
\ee
we have
\be
\label{dL1}
dL_{\ep}(q_{\ep}^i, \dot q_{\ep}^i)\mid_{\ep^k=0}=E(q^i, \dot q^i)
+\frac {d} {dt} \theta.
\ee
From the definitions (\ref{E1}),  (\ref {t1}) and the equation (\ref{dL1}),
it is easy to verify the following important issues.

First, the null Euler-Lagrange 1-form is corresponding to the Euler-Lagrange
equation and the null Euler-Lagrange 1-form is a special case of the coboundary Euler-Lagrange
 1-forms
\be
 E(q^i, \dot q^j)=d\a (q^i, \dot q^j),
\ee
where $\a (q^i, \dot q^j)$ is an arbitrary smooth function of
$ (q^i, \dot q^j)$ 
in $F(T^*M)$.

Secondly, if the Lagrangian $L$ in (\ref{S1}) changes to $L'$ by
adding certain term
\be L(q^i, \dot q^j) \rightarrow L'(q^i, \dot
q^j) =L(q^i, \dot q^j) +{\cal V} (q^i), \ee where ${\cal V} (q^i)
$ is an arbitrary function of $q^i$,  the equation (\ref{dL1})
changes to \be\label{dL2} dL'_{\ep}(q_{\ep}^i, \dot
q_{\ep}^i)\mid_{\ep^k=0}=E'(q^i, \dot q^i) +\frac {d} {dt}
{\theta}, \ee where $E'(q^i, \dot q^i)$  differes from $E(q^i,
\dot q^i)$  by changing $L$ to $L'$ in the expressions, while
$\theta$ has not been changed because ${\cal V}$ does not depend
on $\dot q^i$. In fact, the Euler-Lagrange equation has been
changed by adding a potential-like term that does not depend on
$\dot q^i$. This means that even if by adding a coboundary term ,
the Euler-Lagrange equation does change and canonical form may be
still the same as before.

Thirdly, the Euler-Lagrange 1-forms are not coboundary in general
since $\theta$ is not a coboundary. Therefore, we arrive at an
important theorem for the classical Lagrangian mechanism
\c{GLW01}\c{hyg1} as follows.

{\it Theorem 2.1}

There exists a nontrivial Euler-Lagrange cohomology in the classical
Lagrangian mechanism:

{\centerline {$H_{CM}$:=\{ closed Euler-Lagrange forms\}/\{ exact Euler-Lagrange forms\}.}

Furthermore, owing to the nilpotency of $d$ with respect to
$\ep^k$ in the cotangent space of the congruence on $F(T^*M)$,
$$
d^2L_{\ep}(q_{\ep}^i, {\dot q_{\ep}^j})\mid_{\ep^k=0}=0,
$$
it follows that
 \be
dE(q^i, \dot q^i) +\frac {d} {dt} \om_L =0,
\ee
where  $\om_L$ is the symplectic structure in the Lagrangian
formalism defined by
\be \omega_L = d\theta = \frac {\pa^2 L} {\pa
{\dot q^i} {\pa q^j}} dq^i \we dq^j +\frac {\pa^2 L} {\pa {\dot
q^i}{\pa {\dot q^j}}} dq^i \we d{\dot q}^j.
\ee
And it does not change if the canonical 1-form transforms as
\be
{\theta} \rightarrow {\theta}'={\theta}+d\b(q^i, \dot q^i),
\ee
where $\b(q^i, \dot q^i)$ is an arbitrary function of $(q^i, \dot q^i)$.

It is more important now that another important theorem in the classical
Lagrangian mechanism \c{GLW01} \c{hyg1} can
straightforwardly be established.

{\it Theorem 2.2}:

The symplectic structure $\om_L$ exists and its preserving
property, i.e. the symplectic conservation law with respect to $t$
\be \frac {d} {dt} \omega_L = 0 \ee holds {\it if and only if }
the Euler-Lagrange 1-form is closed with respect to $d$  in the function space $F(T^*M)$, \be d
E(q^i, \dot q^j)=0. \ee

The above closed condition has been named the (closed)
Euler-Lagrange condition in \c{GLW01}\c{hyg1}.

The importance of the theorem  2.2 is to indicate that the
preserving  property of the symplectic structure $\omega_L $  that
always exists in the classical Lagrangian mechanism is directly
related to the Euler-Lagrange cohomology, especially to the closed
Euler-Lagrange condition. On the other hand, the nontriviality of
the Euler-Lagrange cohomology is given by the symplectic structure
and its preserving property. It is simple to see from the
cohomological point of view that although the null Euler-Lagrange
1-form, the coboundary Euler-Lagrange 1-forms satisfy the
Euler-Lagrange condition, it does not mean that the closed
Euler-Lagrange 1-forms can always be exact. Namely, as was
mentioned above, it is guaranteed by the nontriviality of the
Euler-Lagrange cohomology. This  also indicates that $q^i(t)$,
$(i=1, \cdots, n)$, in the Euler-Lagrange condition are {\it NOT}
in the solution space of the Euler-Lagrange equation only. In
fact,
 they are still in the function space associated with the Euler-Lagrange condition in general.
Therefore, as the theorem 2.2 claimed,
the symplectic structure preserving property, i.e. the symplectic 2-form
$\omega_L$  is conserved, holds
not only in the solution space of the equation but also in the
function space in general with respect to the duration of $t$ if and
only if the closed Euler-Lagrange condition is satisfied.

On the other hand, it is interesting to see that if we introduce a
2-form \be {\Om}(q^i, \dot q^i)=d E(q^i, \dot q^j), \ee
then $\Om$ 
may be viewed as a $U(1)$-like curvature 2-form while the
Euler-Lagrange 1-form  the $U(1)$-like connection 1-form.
Therefore, the closed Euler-Lagrange condition is nothing but the
flat connection condition. Furthermore, if for some reason that
the symplectic conservation law is broken,  the broken
pattern may be described by the curvature 2-form $\Om$.

Finally, it should be pointed out that the Euler-Lagrange
cohomological approach may directly be started from the
Euler-Lagrange equation to define the associated Euler-Lagrange
1-form and release from the solution space, then by taking
exterior derivative of the Euler-Lagrange 1-form the theorem 2.2
may also be established straightforwardly.

\subsection{ Euler-Lagrange
cohomology and symplectic structure preserving  in Hamiltonian formalism}

$\quad $ The cohomological concepts, content and theorems in the last subsection for the
Lagrangian formalism of the classical mechanics can also be well established
 in the Hamiltonian formalism.

In order to transfer to the Hamiltonian formalism, we introduce a
family of conjugate momenta from the family of Lagrangian
$L_{\ep}$ in (\ref{lep1}) \be p_{\ep j}=\frac {\pa L_{\ep}} {\pa
\dot {q}_{\ep} ^j}, \ee and take a Legendre transformation to get
the Hamiltonian function in the family \be H_{\ep} :=H(q_{\ep}
^i,p_{\ep j})=p_{\ep k}  {\dot q}_{\ep} ^k -L(q_{\ep} ^i, {\dot
q_{\ep} }^j). \ee Then the exterior derivative of $H_{\ep} $ may
be taken as the one with respect to the set of free parameters:
\be dH_{\ep} =(\f {\pa H_{\ep} } {\pa {p_{\ep i} }}-\dot q_{\ep}
^i)d p_{\ep i} +(\f {\pa H_{\ep} } {\pa { q_{\ep} ^i}}+\dot p_{\ep
i}) d q_{\ep} ^i
+\dot q_{\ep} ^i dp_{\ep i}-\dot p_{\ep i} dq_{\ep} ^i .
\ee

A pair of the families Euler-Lagrange 1-forms now may be introduced as follows: 
\be \label{HEE1} E_{\ep 1} (q_{\ep} ^i, p_{\ep j} )=( \dot
{p}_{\ep j}  + \frac {\pa H_{\ep} } {\pa q_{\ep} ^j})dq_{\ep}
^j,\quad E_{\ep 2} (q_{\ep} ^i, p_{\ep j} )=(
 \frac {\pa H_{\ep} } {\pa p_{\ep j}} -\dot {q}_{\ep} ^j)dp_{\ep j}.
\ee
We may introduce a family of $z_{\ep^l} ^T=(p_{\ep^l} ^T, q_{\ep^L} ^T), p_{\ep^l} ^T
=(p_{\ep ^l 1} ,\cdots, p_{\ep ^l n} ), q_{\ep^l} ^T=(q_{\ep^l} ^1,\cdots,
q_{\ep^l} ^n)$  defined by
\be z_{\ep^l}(t) :=z(t)+\d
z_{\ep^l}(t)=z(t)+{\ep^l} \d z_l(t), \ee where
$$
\d z_{\ep^l}(t)=\f d {d\ep^l}\mid_{\ep^l=0}z_{\ep^l}(t)= \d
z_l(t)$$ is an infinitesimal variation of $z(t)$ along the
direction $l$ in the configuration space. Then the Euler-Lagrange
1-forms in (\ref{HEE1}) become \be \label{HEE2} E_{\ep} (z_{\ep}
, \dot z_{\ep} )=dz_{\ep} ^T(\nabla_{z_{\ep} } H_{\ep}-J \dot
z_{\ep}), \ee where $J$ is a symplectic matrix.

It is straightforward to verify  the following issues in
Hamiltonian formalism:

First, the null Euler-Lagrange 1-forms by setting ${\ep} ^l=0$ give rise to
a pair of the canonical equations as follows
\be
\label{HE1}
\dot q^i=\frac {\pa H} {\pa {p_i}},\qquad
\dot p_j=-\frac {\pa H} {\pa { q^j}}.
\ee
It is easy to see that the second equation comes from the definition of $p_j$ and Legendre
transformation with ${\ep} ^l=0$ while
the first equation is associated with the Euler-Lagrange equation.
In terms of $z$,
we have
\be
\label{HE2}
\dot z=J^{-1}\nabla_z H.
\ee

Secondly, the null forms are the special case of the
coboundary Euler-Lagrange 1-forms, say,
\be
E(z, \dot z)=d\a(z, \dot z),
\ee
where $\a(z, \dot z)$ is an arbitrary function of $(z, \dot z)$.

Thirdly,  from either the expression of $dH_{\ep} $ or 
the definitions of the Euler-Lagrange 1-forms it is easy to see
that the Euler-Lagrange 1-forms are not exact in general.
Therefore, the Euler-Lagrange cohomology in the Hamiltonian
formalism is also nontrivial as it should be .

Fourthly, by taking $d^2 H_{\ep} =0$ and setting ${\ep} ^k=0$
afterwards, it is straightforward to get \be dE(z, \dot z)+\f d
{dt} \om_H =0, \ee where $\om_H $ is the symplectic structure
$\om_L$ transferred into the Hamiltonian formalism \be \om_H =\f 1
2  dz^T \wedge Jdz. \ee Therefore, we have arrived at the
following theorem in the Hamiltonian mechanism.

{\it Theorem 2.3}

In the Hamiltonian mechanism the symplectic structure $\om_H $ preserving law
\be
\f d {dt} \om_H  =0
\ee
holds {\it if and only if} the Euler-Lagrange form is closed:
\be
dE(z, \dot z)=0, \quad  i.e. \quad d(E_1(q^i, p_j)+E_2(q^i, p_j))=0.
\ee

Finally, it should be mentioned that as in the Lagrangian
formalism the Euler-Lagrange cohomological scenario in the
Hamiltonian formalism may also be performed in two slightly
different processes. Namely, it may either start from the exterior
derivative of the Hamiltonian or begin with the canonical
equations. In the second process, the families of the
Euler-Lagrange 1-forms (\ref {HEE1}) and (\ref{HEE2}) may be
introduced directly from the canonical equations (\ref{HE1}) and
(\ref{HE2}) respectively. Then by taking the exterior derivative
of the families of the Euler-Lagrange 1-forms and setting the free parameters being vanish
it also follows the
theorem 2.3.

\section{Euler-Lagrange
Cohomology and Multisymplectic Structure Preserving
in Classical Field Theory}

$\quad $
We now consider the multisymplectic structure and its preserving property
in classical
field theory for a set of generic fields. Most important issue is about the
Euler-Lagrange cohomology and its relation with the multisymplectic
structure preserving property.  We first consider the Lagrangian
formalism and then transfer to the Hamiltonian formalism.

\subsection{Euler-Lagrange
cohomology and multisymplectic structure preserving for classical
field theory in Lagrangian formalism}

$\quad $
For the sake of simplicity, let 
$X^{(1,n-1)}$ be an $n$-dimensional Minkowskian space as base manifold with
coordinates $x^{\mu}$,
$(\mu =0, \cdots,n-1 )$, $M$ the configuration space
 on $X^{(1,n-1)}$ with a set of generic fields $u^i(x)$, $(i=1, \cdots, s)$,
$TM$ the tangent bundle of $M$ with coordinates $(u^i, u_{\mu}^j)$, where
 $u_{\mu}^j=\frac {\pa u^i} {\pa x^{\mu}}$,
  $F(TM)$ the function space on $TM$ etc. We also assume
these fields to be  free of constraints.

The Lagrangian of the fields now is a functional of the set of
generic fields under consideration: \be L(u^i, \dot u^i)=\int
d^{n-1}x {\cal L}(u^i({\bf x},t), u_{\mu}^j({\bf x}, t)),\qquad
u^i(x)=u^i({\bf x},t), \quad etc., \ee and the action is given by
\be S(u^i, u_{\mu}^i)=\int dt L( u^i, \dot u^i)=\int d^nx {\cal
L}(u^i, u_{\mu}^i), \ee where ${\cal L}(u^i({\bf x},t),
u_{\mu}^i({\bf x}, t))$ is the Lagrangian density.

In order to apply Hamilton's principle we first consider how to define the
variation of the action functional $S(u^i(x), u_{\mu}^i(x))$ in a manner analog to the
case of classical mechanics.
In order to achieve this purpose, let us suppose that both
$u^i(x)$ and ${u_{\mu}^i(x)}$ variate by an
infinitesimal increments such that
at a spacetime point of $x$  
the infinitesimal variations of $u^i$ and ${u_{\mu}^i}$  can be
described as follows \be u_{\ep}^i({\bf x},t)=u^i({\bf
x},t)+{\ep^k} \d u_k^i({\bf x},t), \qquad {u_{\mu \ep}^j}({\bf
x},t)=u_{\mu}^j({\bf x},t)+{\ep^k}{\d {u_{\mu k} ^j}({\bf x},t)},
\ee where ${\ep^k}$ are $s$ free parameters  along the $k$-th
direction in the configuration space, and \be \d
u_{\ep^k}^i(x):=\f d {d \ep^l}\mid_{\ep^l=0} u_{\ep}^i(x)=\d
u_k^i(x),\qquad \d u_{\mu \ep^k}^j(x):=\f d {d
\ep^l}\mid_{\ep^l=0}u_{\mu \ep}^j(x)= \d {u_{\mu k} ^j(x)}, \ee
the infinitesimal increments
 of $u^i({\bf x},t)$  and $u_{\mu}^j({\bf x},t)$ at the spacetime point $x$
 respectively. Then the exterior derivatives of
 $u^i({\bf x},t)$  and $u_{\mu}^j({\bf x},t)$ at the spacetime point
 $x$ may be defined as:
\be
 du_{\ep}^i:= \f {\pa u_{\ep}^i} {\pa {\ep^l} }  d{\ep^l}
=d{\ep}^k \d u_k^i,\qquad
d{u_{\mu \ep}^j}:=
\f {\pa u_{\mu \ep}^i} {\pa {\ep^l} }  d{\ep^l} =d{\ep}^k \d u_{\mu k}^i.%
\ee
It should be noticed that 
$du_{\ep}^i$ and $d u_{\mu \ep}^i$
can be regarded as 1-forms with respect to $d\ep^k$.

Now, the Lagrangian becomes a family of Lagrangian functionals
\be\label{Lft} L_{\ep}(u_{\ep}^i, \dot u_{\ep}^i )= \int d^{n-1}x
{\cal L}(u_{\ep}^i({\bf x},t), {u_{\mu \ep}^i({\bf x},t)}), \ee
and the action $S(u^i(x), u_{\mu}^i(x))$ also becomes a family of
functionals \be S_{\ep}=S(u_{\ep}^i(x), u_{\mu \ep}^i(x)). \ee
Then the variation of the action can be defined as its
differentiation with respect to $\ep^l$ and setting $\ep^l=0$
afterwards. Namely, \be \d S:=\f d {d \ep^l}\mid_{\ep^l=0}
S_{\ep}. \ee

Manipulating the variation of the action functional in this manner
and integrating by parts, it  follows that \be \d S_{\ep} =\int
d^nx\{(\f {\pa \cal L_{\ep}} {\pa u_{\ep}^i}
-\pa_{\mu}({ \f {\pa \cal L_{\ep}} {\pa u_{\mu \ep}^i}}))
\d u_{\ep}^i +\pa_{\mu}(\f {\pa \cal L} {\pa u_{\mu \ep}^i} \d
u_{\ep}^i)\}. \ee Assuming $ \d u_{\ep}^i\mid_{\pm \infty}=0, $
and requiring $ \d S_{\ep}=0$ according to Hamilton's principle,
then the Euler-Lagrange equation follows: \be \f {\pa \cal L} {\pa
u^i} -\pa_{\mu}({ \f {\pa \cal L} {\pa u_{\mu}^i}})=0. \ee

On the other hand, the differentiation of the action functional
with respect to the free parameters $\ep^l$ may be given by
\be\label{dS2} d S_{\ep} =\int d^nx\{(\f {\pa \cal L_{\ep}} {\pa
u_{\ep}^i} -\pa_{\mu}({ \f {\pa \cal L_{\ep}} {\pa u_{\mu
\ep}^i}})) d u_{\ep}^i +\pa_{\mu}(\f {\pa \cal L} {\pa u_{\mu
\ep}^i} d u_{\ep}^i)\}.
 \ee It is important to notice that the
integrant in the equation (\ref{dS2}) reads: \be \label{dL3}
d{\cal L}_{\ep}=(\f {\pa \cal L_{\ep}} {\pa u_{\ep}^i}
-\pa_{\mu}({ \f {\pa \cal L_{\ep}} {\pa u_{\mu \ep}^i}}))
du_{\ep}^i +\pa_{\mu}(\f {\pa \cal L} {\pa u_{\mu \ep}^i}
du_{\ep}^i). \ee Let us define a family of the Euler-Lagrange
1-forms \be E_{\ep}( u_{\ep}^i, u_{\mu \ep}^i):= (\f {\pa {\cal
L}_{\ep}} {\pa u_{\ep}^i}
-\pa_{\mu}({ \f {\pa \cal L_{\ep}} {\pa u_{\mu \ep}^i}}))
du_{\ep}^i, \ee and $n$ families of 1-forms that each family
corresponds to a set of canonical 1-forms \be
\theta_{\ep}^{\mu}:={ \f {\pa \cal L_{\ep}} {\pa u_{\mu
\ep}^i}}du_{\ep}^i. \ee Then the equation (\ref{dL3}) becomes \be
\label{dL4} d{\cal L}_{\ep}=E_{\ep}( u_{\ep}^i, u_{\mu
\ep}^i)+\pa_{\mu}\theta_{\ep}^{\mu}. \ee

 Now, similar to the last
section, it is easy to verify the following issues:

First, the null Euler-Lagrange 1-form with $\ep^l=0$, i.e.
\be
E_{\ep}( u_{\ep}^i, u_{\mu \ep}^i)\mid_{\ep^l=0}=0
\ee
gives rise to the Euler-Lagrange equation.

Secondly, $E_{\ep}( u_{\ep}^i, u_{\mu \ep}^i)=0$ is a special case of the coboundary
Euler-Lagrange 1-forms
\be
E_{\ep}( u_{\ep}^i, u_{\mu \ep}^i)=d\a_{\ep}( u_{\ep}^i, u_{\mu \ep}^i),
\ee
where $\a_{\ep}( u_{\ep}^i, u_{\mu \ep}^i)$ a family of arbitrary functions
of $( u_{\ep}^i, u_{\mu \ep}^i)$.  Although they are cohomologically trivial but it can
already be seen that in the Euler-Lagrange 1-forms, $(u^i, u_{\mu}^j)$ are already
{\it NOT} in the solution space of the Euler-Lagrange equation only rather they are
in the function space with the closed Euler-Lagrange condition (see below)
in general.

Thirdly, if the Lagrangian density $\cal L$ in (\ref{Lft}) changes
to $\cal L'$ by adding certain term \be {\cal L}(u^i,  u_{\mu}^j)
\rightarrow {\cal L}'(u^i, u_{\mu}^j)  = {\cal L}(u^i,  u_{\mu}^j)
+{\cal V} (u^i), \ee where ${\cal V} (q^i) $ is an arbitrary
function of $u^i$, the equation (\ref{dL3}) changes to
\be\label{dL4} d{\cal L}'_{\ep}\mid_{\ep^k=0}=E'(u^i, u_{\mu}^i)
+\frac {\pa} {\pa x^{\mu}} {\theta^{\mu}}, \ee where $E'(u^i,
u_{\mu}^i)$  differs from $E(u^i, u_{\mu}^i)$  by changing ${\cal
L}$ to ${\cal L}'$ in the expressions, while a set of $n$
canonical 1-forms $\theta^{\mu}$ have not been changed because
${\cal V}(u^i)$ does not depend on $u_{\mu}^i$. In fact, the
Euler-Lagrange equation has been changed by adding a
potential-like term that does not depend on $u_{\mu}^i$. This
means that even if by adding a coboundary
 term,  the Euler-Lagrange equation does change and the set of canonical forms may  still be the same as before.

Fourthly,  from the equation (\ref{dL4}) it is easy to see that
$E_{\ep}( u_{\ep}^i, u_{\mu \ep}^i)$ in general are not
cohomologically trivial because the families of canonical 1-forms
are not trivial. Therefore, the following theorem  can be
established \c{GLW01} \c{hyg1}.

{\it Theorem 3.1}:

There exists a nontrivial Euler-Lagrange cohomology in the classical
Lagrangian field theory for the set of generic fields $u^i(x)$:

{\centerline {$H_{CFT}$:=\{ closed Euler-Lagrange forms\}/\{ exact Euler-Lagrange forms\}.}}

Furthermore, due to the nilpotency of $d$ with respect to $\ep^k$,
 taking the second exterior derivative of ${\cal L}_{\ep}(u_{\ep}^i, {u_{\mu \ep}^i})$
 and setting $\ep^k=0$
afterwards
$$
d^2{\cal L}_{\ep}(u_{\ep}^i, {u_{\mu \ep}^i})\mid_{\ep^k=0}=0,
$$
it follows that \be dE(u^i, u_{\mu}^i) +\frac {\pa} {\pa
{x^{\mu}}} \om^{\mu} =0, \ee where  $\om^{\mu}$ are $n$ symplectic
structures defined by \be \omega^{\mu} = d\theta^{\mu} = \frac
{\pa^2 \cal L} {\pa { u_{\mu}^i} {\pa u^j}} du^j \we du^i +\frac
{\pa^2 \cal L} {\pa { u_{\mu}^i}{\pa {u_{\nu}^j}}} du_{\nu}^j \we
du^i. \ee And they do not change if the set of $n$ canonical
1-forms transform as \be {\theta^{\mu}} \rightarrow
{\theta^{\mu}}'={\theta^{\mu}}+d\b(u^i, u_{\mu}^i), \ee where
$\b(u^i, u_{\mu}^i)$ is an arbitrary function of $(u^i,
u_{\mu}^i)$.

Then it is now straightforward to get another
important theorem in the classical Lagrangian  field theory \c{GLW01} \c{hyg1}.

{\it Theorem 3.2}:

There exists a set of $n$ symplectic structures  $\om^{\mu}$  and  the  multisymplectic  preserving property,
 i.e. the conservation or divergence free  law of the multisymplectic structures
\be \frac {\pa} {\pa {x^{\mu}} }\om^{\mu} =0 \ee holds {\it if and
only if } the Euler-Lagrange 1-form is closed \be d E(u^i,
u_{\mu}^j)=0. \ee

Similar to the finite dimensional case, it is interesting to see
that if we introduce a new 2-form \be {\Om}(u^i, u_{\mu}^i)=d
E(u^i, u_{\mu}^j). \ee It is easy to see that $\Om$  may be viewed
as a $U(1)$-like curvature 2-form while the Euler-Lagrange 1-form
the $U(1)$-like connection 1-form. Therefore, the closed
Euler-Lagrange condition is nothing but the flat connection
condition. On the other hand, if for some reason that the
multisymplectic conservation law is broken then the broken pattern
may be described by the curvature 2-form $\Om$.

It is also important to notice that the multisymplectic structure
preserving  property is directly linked with the closed
Euler-Lagrange condition. And although the null Euler-Lagrange
1-form, the coboundary Euler-Lagrange 1-forms satisfy the
Euler-Lagrange condition, it does not mean that the closed
Euler-Lagrange 1-forms can always be exact as was pointed out
above. In addition, $u^i(x)$'s in the Euler-Lagrange condition are
{\it NOT} in the solution space of the Euler-Lagrange equation
only in general. Therefore, the  multisymplectic structure
preserving  property, i.e. the conservation law of the set of $n$
 symplectic 2-forms
$\omega^{\mu}$, holds not only in the solution
space of the equation but also in the function space with the
closed Euler-Lagrange condition in general.

\subsection{Euler-Lagrange
cohomology and multisymplectic structure preserving
for classical field theory in Hamiltonian formalism}

$\quad $ The most concepts, content and theorems in the last
subsection for the Lagrangian field theory of a set of the classical generic
fields can also be well established  in the Hamiltonian formalism.
In order to apply the Euler-Lagrange cohomological approach to
Hamiltonian formalism for the classical field theory with a set of
generic fields, we first have to define a "momentum" that is
canonically conjugate to the field variables
\be \pi_j(x)=\frac
{\pa {\cal L}} {\pa \dot {u}^j},
\ee
and take a Legendre
transformation to get the Hamiltonian density
\be
{\cal H}(u^i,\pi_j)=\pi_k(x) {\dot u}^k(x) - {\cal L}(u^i, {\dot u}^j).
\ee
The Hamiltonian then is given by
\be
H(t)=\int d^{n-1}x {\cal
H}(x),
\ee
with the Legendre transformation
\be
H(t)=\int
d^{n-1}x\pi_k(x) {\dot u}^k(x) - L(t). \ee

From the Legendre transformation, it follows one  of a pair field
equations in canonical formalism \be \label{HEQ1}{\dot u}^k(x)=\f
{\pa \cal H} {\pa {\pi_k (x)}},
 \ee
  \be \label{HEQ2}\dot {\pi}_k(x)=-\f {\pa \cal H} {\pa
{u^k (x)}} +\nabla_a {\f {\pa \cal H} {\pa ({\nabla_a {u^k
(x)}})}},\quad a=1,\cdots,n-1.
 \ee
The second equation (\ref{HEQ2}) comes from the Euler-Lagrange
equation.

Now we may introduce a pair of the Euler-Lagrange 1-forms
associated with the pair of the canonical equations in the
Hamiltonian formalism \be E_1=d\pi_k\{\f {\pa \cal H} {\pa {\pi_k
(x)}}-{\dot u}^k(x)\},\quad E_2=du^k \{ \dot {\pi}_k(x)+\f {\pa
\cal H} {\pa {u^k (x)}} -\nabla_a {\f {\pa \cal H} {\pa ({\nabla_a
{u^k (x)}})}} \}. \ee It is straightforward to prove the following
formula: \be d(E_1+E_2)=\pa_t{\om^0}-\nabla_a{\om^a}, \ee where
${\om^0}$ and ${\om^a},   a=1,\cdots,n-1,$ a set of $n$ symplectic
2-forms \be \om^0=d{\pi_k} \we du^k,\qquad \om^a= d(\f {\pa {\cal
H}} {\pa ({\nabla_a {u^k (x)}})}) \we du^k. \ee Now we may
establish the following theorem for the multisymplectic structure
conservation law in the Hamiltonian formalism for classical field
theory with a set of generic fields.

{\it Theorem 3.3}:

In the Hamiltonian formalism for classical field theory of a set
of generic fields $u^i(x)$ on $n$-dimensional Minkowskian spacetime there
exists a set of $n$ symplectic 2-forms
 ${\om^0}$  and ${\om^a},   a=1,\cdots,n-1,$  and the multisymplectic structure
preserving law, i.e. their conservation law,
\be
\pa_t{\om^0}-\nabla_a{\om^a}=0,\qquad a=1,\cdots,n-1,
\ee holds {\it if and only if}
$E_1+E_2$ is closed. Namely,
\be d(E_1+E_2)=0.
\ee

It should be noticed that here the exterior derivatives of $E$s,
i.e. $dE$s, have been taken and it should be understood as the
exterior calculus in the generic function space such that $d:
\Om^k \rightarrow \Om^{k+1}$ with $d^2=0$ ( see, for example,
\c{AC94} ). On the other hand, it may be more directly regarded as
the exterior differentiation in the configuration space in the
sense of last subsection. Namely, introducing a set of free
parameters $\ep^l, l=1,\cdots, s$, such that each  variable and
functional etc becomes a family of relevant objects with respect
to the free parameters, then the exterior differentiation in the
configuration space may be manipulated as the one with respect to
the free
parameters $\ep^l$ and finally, after all calculations are completed,
 setting $\ep^l=0$. 

It should also be pointed out that in the above content of this
subsection, the Euler-Lagrange cohomological approach starts from
the canonical equations and the introduction of the Euler-Lagrange
1-form(s) is associated with the canonical equations, while the
symplectic structure and its conservation law are derived from the
exterior derivative of the Euler-Lagrange 1-forms. The entire
scenario seems, at least superficially,  to be nothing related
with the exterior derivative of
 Hamiltonian. However, the
scenario may also begin with the  exterior derivative of  Hamiltonian that is
analog with the case in the Hamiltonian
mechanism. Let us describe this issue in what follows.

In order to more directly take the  exterior derivative of  Hamiltonian 
 we introduce a family of Hamiltonian functionals
\be
H_{\ep}(t)=\int d^{n-1}x {\cal H}_{\ep}(x).
\ee
and make the Legendre transformation in the family as well
\be
{\cal H}_{\ep}(u_{\ep}^i,\pi_{\ep j})=\pi_{\ep k}(x) {\dot u_{\ep}}^k(x)
- {\cal L}_{\ep}(u_{\ep}^i, {\dot u}_{\ep}^j).
\ee
Then  the  exterior derivatives of Hamiltonian may be taken as the differentiation of $H_{\ep}(t)$ with respect to $\ep$ as follows
$$
dH_{\ep}(t) =\int d^{n-1} x  d{\cal H}_{\ep} (u_{\ep}^i,\pi_{\ep
j})
$$$$
=\int d^{n-1} x\{\f {\pa {\cal H}_{\ep} } {\pa \pi_{\ep j} } d\pi_{\ep j}
 +\f {\pa {\cal H}_{\ep} } {\pa u_{\ep}^i } du_{\ep}^i
-{\n_a}(\f {\pa {\cal H}_{\ep} } {\pa (\n_a u_{\ep}^i
)})du_{\ep}^i +\n_a (\f {\pa {\cal H}_{\ep} } {\pa(\n_a u_{\ep}^i
)} du_{\ep}^i ) \}.
$$
Now we introduce a pair of families of the Euler-Lagrange 1-forms
\be E_{\ep 1} =d\pi_{\ep k} \{\f {\pa \cal H_{\ep} } {\pa
{\pi_{\ep k}  (x)}}-{\dot u_{\ep} }^k(x)\},\quad E_{\ep 2}
=du_{\ep} ^k \{ \dot {\pi}_{\ep k} (x)+\f {\pa \cal H_{\ep} } {\pa
{u_{\ep} ^k (x)}} -\nabla_a {\f {\pa \cal H_{\ep} } {\pa
({\nabla_a {u_{\ep} ^k (x)}})}} \}, \ee and \be \theta_{\ep} ^0=\f
{\pa} {\pa t}u_{\ep} ^l d\pi_{\ep l}-\f {\pa \pi_{\ep l} } {\pa t}
du_{\ep} ^l, \qquad \theta_{\ep} ^a=\n_a(\f {\pa{\cal H_{\ep} }}
{\pa (\n_a {u_{\ep} ^k})} du_{\ep} ^k). \ee
Hence
 $d H_{\ep}(t)$ becomes
\be dH_{\ep}(t) =\int d^{n-1} x \{E_{\ep 1}+ E_{\ep 2}+
\theta_{\ep} ^0+\theta_{\ep} ^a\}.
\ee Taking the second exterior
derivative of $H(t)$, due to the nilpotency of $d$, it follows
that \be d(E_{ \ep 1}+ E_{\ep 2}) +\f {\pa} {\pa t} \om_{\ep}
^0-\n_a\om_{\ep} ^a=0, \ee where \be \om_{\ep} ^0 =du_{\ep} ^k \we
d\pi_{\ep k}  \qquad \om_{\ep} ^a=du_{\ep} ^k \we d(\f {\pa {\cal
H}_{\ep} } {\pa (\n_a u_{\ep} ^k)}). \ee By setting ${\ep^l}=0$,
we re-derive the theorem 3.3.

\section{Cohomological Approach to  Hamiltonian-like ODEs, PDEs
and Their  Symplectic and Multisymplectic Properties }

$\quad$
As was mentioned above, it is worthwhile to notice that 
in the previous sections there are some slight differences between
two Euler-Lagrange cohomological approaches in both Lagrangian and
Hamiltonian formalism for  classical mechanism and classical field
theory.

In one of the approaches the cohomological scenario  starts from
the variation of the action functional in Lagrangian formalism or
the exterior derivative of the Hamiltonian functional in
Hamiltonian formalism so that the exterior derivative of the
Lagrangian $dL$ or that of the Hamiltonian $dH$ in the function
space gives rise to the relation between the Euler-Lagrange form
and the divergence of the canonical 1-form(s) that exhibits the
nontriviality of the Euler-Lagrange forms and the cohomology. Then
$d^2L=0$ or $d^2H=0$ leads to the theorem on that the closed
Euler-Lagrange condition is the necessary and sufficient condition
for the symplectic and multisymplectic structure preserving laws
in classical mechanism and field theory respectively.

In another approach, however, the Euler-Lagrange cohomological
scenario may directly begin with the definition of  the
Euler-Lagrange 1-form(s) associated with the Euler-Lagrange
equation or the canonical equations. The nontriviality of the
Euler-Lagrange cohomology originates from the principal parts of
the Euler-Lagrange equation or the canonical equations that lead
to the one(s) in the Euler-Lagrange 1-form(s) that are essentially
same as the canonical 1-form(s) in each case. Then the exterior
derivatives of the Euler-Lagrange 1-forms establish the relation
between the closed Euler-Lagrange condition and the conservation
laws of the symplectic and the multisymplectic structures in the
classical mechanism and field theory respectively.

As a matter of fact, although
 the essentials of the two approaches are almost the same, these slight differences however also
indicate that for certain types of given ODEs and PDEs no matter
whether  the associated Lagrangian and/or Hamiltonian exist or
not, the cohomological scenario may directly be applied to them by
starting with the introduction of the relevant what will be called
the Euler-Lagrange-like 1-forms or symplecticity 1-forms, and
progressing further to see whether there exist what may be called
the Euler-Lagrange-like, or symplectic cohomology,
symplectic/multisymplectic
 structures and their preserving laws.
In other wards, the Euler-Lagrange-like cohomological, or the symplectic cohomological,  scenario
may directly be applied to certain types of ODEs and PDEs that may be named the
Hamiltonian-like
ODEs and PDEs respectively.

In what follows, we first explore how to apply the cohomological
approach to the type of Hamiltonian-like ODEs in subsection 4.1.
Then we deal with the type of Hamiltonian-like PDEs in subsection
4.2. We show that so-called the type of Hamiltonian PDEs
introduced in \c{TB97} is just a case of   Hamiltonian-like PDEs.
We also show that in each case there always exist
Lagrangian/Hamiltonian-like functional for the Hamiltonian-like
ODEs and PDEs respectively such that the vanishing variational of
the action-like functional that is integral of the relevant
Lagrangian-like functional may always lead to the corresponding
Hamiltonian-like ODEs or PDEs.

\subsection{Symplectic
cohomological approach to  Hamiltonian-like ODEs and their symplectic
structure preserving law}

$\quad$ In this subsection, we explore this aspect and show that
the Euler-Lagrange-like, or symplectic cohomological approach is
available for certain types of  Hamiltonian-like ODEs. We also
furnish these ODEs with what is called the Lagrangian-like  or the
Hamiltonian-like functional and show that these ODEs may be
derived from what may be called Hamilton-like's principle for an
action-like functional that is given by the integral of the
Lagrangian-like functional.

We first consider a type of ODEs that may be viewed as the variety of a pair of the canonical equations
(\ref{HE1})  in classical mechanics as follows
\be
\label{HE1'}
\dot q^i=\frac {\pa {\cal H}} {\pa {p_i}},\qquad
\dot p_j=-\frac {\pa {\cal H}} {\pa { q^j}},
\ee
 where ${\cal H}$ differs with $H$ by an arbitrary smooth enough function of
 $p_j, q^i$ on the phase space. Then it is straightforward
to verify that the same cohomology associated with the original Hamiltonian system also available to this
case.

Let us further consider the following type of ODEs that is a mimic of (\ref{HE2}):
\begin{equation}
\label{ODE}
J{\dot z} = \nabla_zS(z), 
\end{equation}
where  $J$ a symplectic matrix 
$z(t)$ the dependent variables no matter whether it is the
canonical variables on the phase space, $S(z)$  an arbitrary
smooth enough function of $z(t)$ no matter whether it is a
Hamiltonian.

It is known from the previous sections that the cohomological
scenario may be begun with introducing the Euler-Lagrange-like
1-form associated to the ODEs. In order to do so, we assume that
the solution space of the ODE exists and release all dependent
variables $z(t)$ from the solution space by some infinitesimal
variations described by a set of free parameters. For example, the
variables $z(t)$ may be changed to \be z(t) \quad \rightarrow
\quad z_{\ep}(t)=z(t)+\ep^l \d z_l, \ee where $\d z_l$ is an
infinitesimal variation of $z$ along the direction $l$ in the
function space, and so on. Then the exterior derivative of $z$ in
the function space may be taken as the one with respect to the
free parameters $\ep^l$ as follows \be dz:=dz_{\ep}=\f {\pa
z_{\ep}} {\pa \ep^l} d\ep^l=d\ep^l \d z_l. \ee In what follows,
all exterior derivatives in the function space are taken in this
sense. And for the sake of simplicity, the omission of the $\ep^l$
will be taken without mentioning.

Now  we are ready to introduce the
Euler-Lagrange-like 1-form associated to
the ODEs:
\begin{equation}
\label{ODEE}
E(z, \dot z  ):=dz^T\{J { \dot z} -\nabla_zS(z) \}.
\end{equation}

It is easy to see that the null Euler-Lagrange-like 1-form gives rise to
 the type of ODEs (\ref{ODE}) and it is a special case of
the coboundary Euler-Lagrange-like 1-forms
\begin{equation}
E(z,  \dot z)=d\a(z,  \dot z ),
\end{equation}
where $\a(z,  \dot z )$ is an arbitrary function of $(z,  \dot z )$.

By taking the exterior derivative $d$ of the Euler-Lagrange-like 1-form, it is
straightforward to prove that
\be
dE(z, \dot z )=\f 1 2 \f d {dt} \om, \qquad \om=dz^T \wedge Jdz.
\ee
where $\omega$ is formally the same as $\om_H$ as it should be if $z$ rely
on the phase space.
This means that the following symplectic structure preserving  equation
\begin{equation}
\f  d  {dt} \omega =0
\end{equation}
holds {\it if and only if} the Euler-Lagrange-like 1-form is
closed:
\begin{equation}
dE(z, \dot z)=0.
\end{equation}

It should also be mentioned that the following issues can  easily
be verified. From the definition of the Euler-Lagrange-like 1-form
(\ref{ODEE}), it follows the nontriviality of the
Euler-Lagrange-like 1-form in general since the first  term in the
definition is the mimic of the canonical 1-form that is obviously
not trivial so that a nontrivial Euler-Lagrange-like cohomology,
or the symplectic cohomology, associated with this type of ODEs
can be introduced

{\centerline {$H_{ODE}$:=\{ closed Euler-Lagrange-like forms\}/\{ exact Euler-Lagrange-like forms\}.}}

In fact, if the variables $z(t)$ are canonical variables, this type of ODEs
(\ref{ODE}) is the same as (\ref{HE1'}) so that they share the same cohomology with
the canonical equation (\ref{HE1}) and (\ref{HE2}). As a matter of fact, as was
pointed, in this case
the Euler-Lagrange-like form (\ref{ODEE}) and the Euler-Lagrange form (\ref{HEE2}) associated with
 the canonical equation  (\ref{HE2})  differ by an exact form.

In addition, similar to all cases in the classical mechanism, the symplectic structure preserving law holds not only in the solution space of the ODEs but also in the
function space associated with the Euler-Lagrange-like cohomology.

Finally, in the case of $z(t)$ is not the canonical coordinates
and momenta, let us consider how to introduce the action-like
functional for this type of ODEs. Regarding the function $S(z)$ as
a Hamiltonian-like function and introducing a Lagrangian-like
functional $L(z, \dot z)$ by 
\be
L(z, \dot z)=\f 1 2 z^T J \dot z -S(z), \ee then a action-like
functional $A(z)$ may be introduced as follows
\be A(z)=\int_a^b
dt L(z, \dot z).
\ee
The variation of the action-like functional
may also manipulated as the differentiation with respect to the
free parameters
$$
\d A(t):=\f d {d \ep^k} A_{\ep}(t)\mid_{\ep^k=0}=\int_a^bdt \f  d {d\ep ^k} L_{\ep}(z_{\ep},
\dot z_{\ep})\mid_{\ep^k=0}.
$$
On the other hand, it is straightforward to calculate that \be dL_{\ep} =dz_{\ep}
^T(Jd \dot z_{\ep} -\n_{z_{\ep} } S_{\ep} (z_{\ep} ))+\f 1 2  \f d
{dt} (z_{\ep} ^TJ dz_{\ep} ). \ee
Then the Hamilton-like's
principle leads to the original ODE (\ref{ODE}) and
$d^2L_{\ep}\mid_{\ep=0}=0$ gives rise to the following wanted
identity: \be dE(z, \dot z)+\f 1 2 \f d {dt} \om=0,\qquad \om=dz^T
\we Jdz. \ee

Furthermore, we even may progress further to some type of ODEs
that mimic the canonical equation(s) in the Hamiltonian mechanism.
For example, the equation (\ref{ODE}) may be generalized to the
following type of Hamiltonian-like ODEs: \be \label{HLE} K\dot
f(t)=\nabla_f S(f), \ee where $K$ is any $n \times n, n=2k+1, k\in
Z_{+} $ antisymmetric nonsingular matrix, $f(t)$ an $n \times 1$
matrix variables, $S(f)$ an arbitrary smooth enough function of
$f$.

The symplectic cohomological scenario may also begin with by
introducing the associated Euler-Lagrange-like 1-form \be
\label{ELL1} E( f, \dot f):=df^T(K\dot f(t)-\nabla_f S(f)). \ee

Taking the exterior derivative of the Euler-Lagrange-like 1-form,
it follows for this type of Hamiltonian-like ODEs that \be dE( f,
\dot f)+\f 1 2 \f d {dt} df^T\we K df=0. \ee Namely,   $\tau=df^T
\we K df$ is a symplectic structure associated with this type of
Hamiltonian-like ODEs and it is conserved {\it if and only if} $
dE( f, \dot f)=0$.

Similarly, the following issues can be verified.
First, the null Euler-Lagrange-like 1-form gives rise to the Hamiltonian-like ODEs (\ref{HLE}) and it is a special case of the
coboundary Euler-Lagrange-like 1-forms.
 Secondly, the definition of ({\ref{ELL1}}) shows that the Euler-Lagrange-like 1-form is nontrivial since the first term in it is
a mimicry of the canonical 1-form in the Hamiltonian mechanism so that there is a nontrivial
Euler-Lagrange-like cohomology, or the symplectic cohomology,
associated to this type of Hamiltonian-like ODEs.
Thirdly, the symplectic structure preserving  equation can be derived here by taking exterior
derivative of the Euler-Lagrange-like 1-form. And it is
not dependant on the solution space of the type of Hamiltonian-like ODEs in general
 but can be applied to their solution space. 
In fact,  the symplectic structure preserving  equation  holds not only in the solution space of the
ODEs but also
in the function space relevant to the Euler-Lagrange-like cohomology as well.
Finally, the action-like functional for this type of ODEs may be found as follows
\be
A(t)=\int_a^b dt L(f, \dot f), \qquad L(f, \dot f)=\f 1 2 f^T K \dot f - S(f).
\ee

\subsection{Symplectic cohomological approach to  Hamiltonian-like PDEs
and their  multisymplectic structure preserving law}

$\quad$ We now consider the application of the cohomological approach to  Hamiltonian-like PDEs
and their  multisymplectic structure preserving law.
 It is worthwhile to emphasize that analog to the case of classical mechanism,
for a kind of
given PDEs no matter the associated Lagrangian and/or Hamiltonian are known or not, the cohomological
scenario
may directly be applied to them
by starting with the introduction of the associated
Euler-Lagrange-like 1-forms and
progressing further to see whether the Euler-Lagrange-like cohomology, multisymplectic structures
and their preserving law exist.
In other wards, the Euler-Lagrange-like cohomological scenario may directly be applied to the PDEs.
Similar to the case of the Hamiltonian-like ODEs, this type of PDEs may be called the Hamiltonian-like
PDEs.

Let us first consider the variety of the Hamiltonian PDEs for
field theory as follows: \be \label{HEQ1'}{\dot u}^k(x)=\f {\pa
\cal H'} {\pa {\pi_k (x)}},
 \ee
  \be \label{HEQ2'}\dot {\pi}_k(x)=-\f {\pa \cal H'} {\pa
{u^k (x)}} +\nabla_a {\f {\pa \cal H'} {\pa ({\nabla_a {u^k
(x)}})}},\quad a=1,\cdots,n-1,
 \ee
where ${\cal H'}$ differ from ${\cal H}$ by an arbitrary function
of ${u^i, \pi^i}$:
 \be
 {\cal H'}={\cal H}+\gamma{(u^i, \pi^i)},
\ee where $\gamma{(u^i, \pi^i)}$ is an arbitrary function of
${(u^i, \pi^i)}$.

Then it is straightforward to verify that the set of PDEs
(\ref{HEQ1'}) and (\ref{HEQ2'}) have the similar Euler-Lagrange
cohomology and share the same multisymplectic conservation law
with the set of canonical equations (\ref{HEQ1}) and (\ref{HEQ2})
for classical field theory in Hamiltonian formalism.

We now consider the following type of so called the Hamiltonian
PDEs introduced first by Bridges through the hypothesis that
adding certain term to canonical equation of motion (an ODE that
is symplectic preserving)  in Hamiltonian mechanism \c{TB97}.  We
will make use of the similar notations in \c{TB97} \c{BR99}
\c{SR00}
\begin{equation}
\label{TB} M {z_{x_1}} + \ep K {z_{x_2}} = \nabla_zS, \qquad  \ep
=\pm 1,
\end{equation}
where $M$ and $K$ are any antisymmetrical matrixes and
$$
{z_{x_i}}=\f  {\pa z}  {\pa {x_i}} , \qquad i=1,2.
$$
Introducing the Euler-Lagrange-like 1-form
\begin{equation}
\label{TBE} E(z, {z_{x_1}} , {z_{x_2}} ):=dz^T\{ M { z_{x_1}} +
\ep K {z_{x_2}} -\nabla_zS \},
\end{equation}
it is easy to see that the null Euler-Lagrange-like 1-form gives rise to
 the type of equations (\ref{TB}) and it is a special case of
the coboundary Euler-Lagrange-like 1-forms
\begin{equation}
E(z,  z_{x_1} , z_{x_2} )=d\a(z,  z_{x_1} , z_{x_2} ),
\end{equation}
where $\a(z,  z_{x_1} , z_{x_2} )$ is an arbitrary function of $(z,  z_{x_1} , z_{x_2} )$.

Now by taking the exterior derivative $d$ of the
Euler-Lagrange-like 1-form, it is straightforward to prove that
\be dE(z, z_{x_1} , z_{x_2} )+\f 1 2 \pa_{x_1} (dz^T \wedge M dz)+
\ep \f 1 2 \pa_{x_2} (dz^T \wedge K dz)=0. \ee This means that the
following multisymplectic structure preserving  equation
\begin{equation}
\pa_{x_1}  \omega + \ep \pa_{x_2}  \tau=0,
\end{equation}
where \be \omega=dz^T \wedge M dz, \quad \tau= dz^T \wedge K dz,
\ee holds {\it if and only if} the Euler-Lagrange-like 1-form is
closed, i.e.
\begin{equation}
dE(z, z_{x_1} , z_{x_2} )=0.
\end{equation}

It should be mentioned that first from the definition of the Euler-Lagrange-like 1-form
(\ref{TBE}), it is not trivial in general since the first two terms in
the definition are the
canonical 1-forms which are obviously not trivial so that a nontrivial  cohomology
associated with this type of Hamiltonian-like PDEs can be introduced

{\centerline {$H_{PDE}$:=\{ closed Euler-Lagrange-like forms\}/\{ exact Euler-Lagrange-like forms\}.}}

Secondly, the multisymplectic structure preserving  equation derived here is
not dependant on the solution space of the type of Hamiltonian-like PDEs in general
 but can be applied to the type of Hamiltonian-like PDEs
so that it is held not only in the solution space of the PDEs but also
in the function space relevant to the cohomology as well.

Thirdly, this type of Hamiltonian-like PDEs can directly be generalized to the higher dimensional space/spacetime  cases.

Fourthly, the action-like functional for this type of PDEs may be
introduced as follows \be
A(t)=\int_a^b dx^2 {\cal L} (z, z_{x_1}, z_{x_2}), 
\ee where the Lagrangian-like functional is given by
\be
{\cal L} (z,
z_{x_1}, z_{x_2})=\f 1 2 z^T( M z_{x_1} + \ep K z_{x_2}) - S(z).
\ee It is straightforward to prove that the vanishing variation of
$A(t)$ leads to the original PDEs (\ref{TBE}).

Finally, it is straightforward to generalize this type of Hamiltonian-like PDEs to higher dimensional case.

\section {Remarks}

$\quad $
A few remarks are in order:

1. The approach presented in the previous paper by Guo, Li and Wu \c {GLW01},
  in the talk \c {hyg1} and in this paper to the
symplectic or multisymplectic geometry for classical mechanism and
classical field theory in the Lagrangian and Hamiltonian
formalisms respectively is more or less different
from other approaches ( see, for example, \c{TB97} \c{MPS98} ). The Euler-Lagrange 
cohomological
concepts and relevant content
such as the Euler-Lagrange 
1-forms, the null Euler-Lagrange 
1-forms, the coboundary Euler-Lagrange 1-forms as well as the
(closed) Euler-Lagrange conditions have been introduced and they
have played very crucial roles in each case to show that the
symplectic or the multisymplectic structure and the relevant
preserving property hold in the function space with the relevant
Euler-Lagrange condition in general rather than in the solution
space of the equation(s) of motion only.
It has been emphasized  that the Euler-Lagrange 
cohomology in each relevant case is
nontrivial and it is very closely related to the symplectic or multisymplectic
structures as well as the relevant preserving property. As a matter of fact, these two aspects, i.e.
the Euler-Lagrange cohomology vs. the symplectic or multisymplectic structure preserving, are two
sides of
the same coin in some sense.

It should pointed out that the content of the Euler-Lagrange
cohomology and the role-played by the cohomology in each case
should be further studied not only in classical level but also in
quantum level as well. And needless to say, this cohomological
scenario should also be generalized to the case with more generic
base manifolds and configuration spaces and applied to whatever
the least variational principle and the Euler-Lagrange equation
are available, such as the mechanics and field theory with
different types of constraints, different types of field theories
and so on so forth.

2. It should be emphasized as was mentioned in the content that
the approach with the Euler-Lagrange cohomological concepts can
also be directly  applied to certain types of Hamiltonian-like
ODEs and PDEs no matter whether they have known Lagrangian and /or
Hamiltonian associated with. And we have found that  if the
cohomological approach works for the Hamiltonian-like ODEs and
PDEs, may also be found some artificial action-like functional,
Lagrangian-like functional and Hamiltonian-like functional
associated with them in certain sense that the Hamilton-like's
principle may applied to get the original ODEs and PDEs. On the
other hand, from the symplectic cohomological point of view, with
in the same cohomology class, those ODEs and PDEs may differ by
some exact forms respectively.

This kind of cohomological approach to ODEs and PDEs should also be payed more attention and to explore further
the content and generalization.
On the other hand, if the cohomological approach works for the Hamiltonian-like ODEs and PDEs, it
is still
interesting to see whether these ODEs and PDEs may be prolonged in certain manner to find
even artificially some more realistic
Lagrangian and/or Hamiltonian functional associated with them.

3. The difference discrete variational principle, difference discretized version of the Euler-Lagrange cohomology,
 symplectic/multisymplectic structure
and the relevant preserving property for the classical mechanism and classical field theory mainly in discrete
Lagrangian formalism have also been studied and have been applied to the symplectic algorithm and
the
multisymplectic algorithm \c{GLW01} \c{hyg1} \c {GLW02} as well as the symplectic/multisymplectic
structure preserving in the finite element method \c{GJLW03} \c{kw1}.  Some new results on these issues have been
established by the present authors. 
We will present our
results elsewhere \c{GLWW01}.

4.  Finally, it should also be emphasized that there exist lots of
other problems to be studied on the Euler-Lagrange cohomology and
related topics not only in mathematics but also in physics. For
instance, the relation between the Euler-Lagrange cohomology and
the second variation of the action functional,
 the relation between  the Euler-Lagrange cohomology and the problems on the existence and stability of the solutions of the
Euler-Lagrange equations, Hamiltonian-like ODEs and PDEs, the
relation between the symplectic group or algebraic symmetry
 cohomology and the Euler-Lagrange cohomology, the symplectic/multisymplectic structure preserving properties,
the physical application of the Euler-Lagrange cohomology,
symplectic and multisymplectic structure conservation laws and so
on so forth.

\vskip 5mm {\parindent=0mm\bf\large Acknowledgements}

Guo, Li and Wu would like to thank Professors and/or Drs J.
Butcher, Y.B. Dai,  J.B. Chen, J.L. Hong, A. Iserles  R.I.
McLachlan, D. Lewis, H. Munthe-Kaas, P. Olver, B.R. Owren, M.Z.
Qin, S. Reich, Z.J. Shang, X.C. Song, G. Sun, Y.F. Tang, L.H.
Wang, S.H. Wang, M. West, Z. Xu for valuable discussions and
comments, specially during the International Workshop on
Structure-Preserving Algorithms, March 25-31, 2001, Beijing. This
work is partly supported by National Science Foundation of China.

\end{document}